\documentclass[aps,prl,twocolumn,showpacs,superscriptaddress,showpacs]{revtex4-1}
\usepackage{xcolor}
\usepackage{physics}
\usepackage{graphicx}
\usepackage{dcolumn}
\usepackage{bm}

\begin{document}

\preprint{APS/123-QED}

\title{Excited quantum Hall effect: enantiomorphic flat bands in a Yin-Yang Kagome lattice}

\author{Yinong Zhou}
\affiliation{%
 Department of Materials Science and Engineering, University of Utah, Salt Lake City, Utah 84112, USA\\
 }%

\author{Gurjyot Sethi}
\affiliation{%
 Department of Materials Science and Engineering, University of Utah, Salt Lake City, Utah 84112, USA\\
 }%

\author{Hang Liu}
\affiliation{%
 Department of Materials Science and Engineering, University of Utah, Salt Lake City, Utah 84112, USA\\
 }%

\author{Zhengfei Wang}
\affiliation{%
	Hefei National Laboratory for Physical Sciences at the Microscale,
	CAS Key Laboratory of Strongly-Coupled Quantum Matter Physics,
	University of Science and Technology of China, Hefei, Anhui 230026, China
}%

\author{Feng Liu}%
\email{fliu@eng.utah.edu}
\affiliation{%
 Department of Materials Science and Engineering, University of Utah, Salt Lake City, Utah 84112, USA\\
 }%

\date{\today}

\begin{abstract}
Quantum Hall effect (QHE) is one of the most fruitful research topics in condensed-matter physics. Ordinarily, the QHE manifests in a ground state with time-reversal symmetry broken by magnetization to carry a quantized chiral edge conductivity around a two-dimensional insulating bulk. We propose a theoretical concept and model of non-equilibrium excited-state QHE (EQHE) without intrinsic magnetization. It arises from circularly polarized photoexcitation between two enantiomorphic flat bands of opposite chirality, each supporting originally a helical topological insulating state hosted in a Yin-Yang Kagome lattice. The chirality of its edge state can be reversed by the handedness of light, instead of the direction of magnetization as in the conventional quantum (anomalous) Hall effect, offering a simple switching mechanism for quantum devices. Implications and realization of EQHE in real materials are discussed.
\end{abstract}

\maketitle

Quantum Hall effect (QHE) displays a quantized Hall conductivity that is only determined by physical constants \cite{klitzing1980new} and fundamentally rooted in topological invariance of electronic states \cite{thouless1982quantized,laughlin1983anomalous,haldane1983nonlinear,haldane1988model}. It was first observed in a two-dimensional (2D) electron gas subjected to a strong magnetic field \cite{klitzing1980new} that splits the parabolic electronic bands into discrete Landau levels (LLs), each characterized with a topological invariant Chern number $C = 1$ \cite{thouless1982quantized,laughlin1983anomalous,haldane1983nonlinear,haldane1988model}.  For an integer QHE, the quantized transverse Hall conductivity equals to the filling of LLs ($n$), $\sigma_{xy} = n\cdot{e^2}/{h}$ [Fig. 1(a)].  Haldane proposed a condensed-matter version of QHE without LLs [the quantum anomalous Hall effect (QAHE)] in a Chern insulator, featured with a topological gap separating a valence band (VB) having a non-zero Chern number (e.g., $C = 1$) from a conduction band (CB) having $C = 0$. It gives rise to a $\sigma_{xy} = C\cdot{e^2}/{h}$ [Fig. 1(b)]. Recently, the quantum spin Hall effect (QSHE) was theoretically proposed by Kane-Mele \cite{kane2005quantum} and Bernevig-Hughes-Zhang \cite{bernevig2006quantum}, featured with a topological gap separating a VB having a non-zero $Z_2$ invariant \cite{kane2005z} or spin Chern number \cite{sheng2006quantum} (e.g., $C^s = -1$) from a CB having $C^s = 0$. It gives rise to a spin Hall conductivity $\sigma^s_{xy} = C^s\cdot{e}/{2\pi}$ [Fig. 1(c)].  The QSHE [or a 2D topological insulator (TI)] contains two copies of QAHE (Chern insulator) so that by breaking the time-reversal symmetry (TRS) of a TI, namely creating a magnetic TI, one obtains a Chern Insulator \cite{liu2008quantum}. Excitingly, both QSHE \cite{konig2007quantum} and QAHE \cite{chang2013experimental} have been experimentally confirmed in real materials. These pioneering works have fostered an ever-growing field of topological physics and materials.

\begin{figure*}
	\includegraphics[width=1.6\columnwidth]{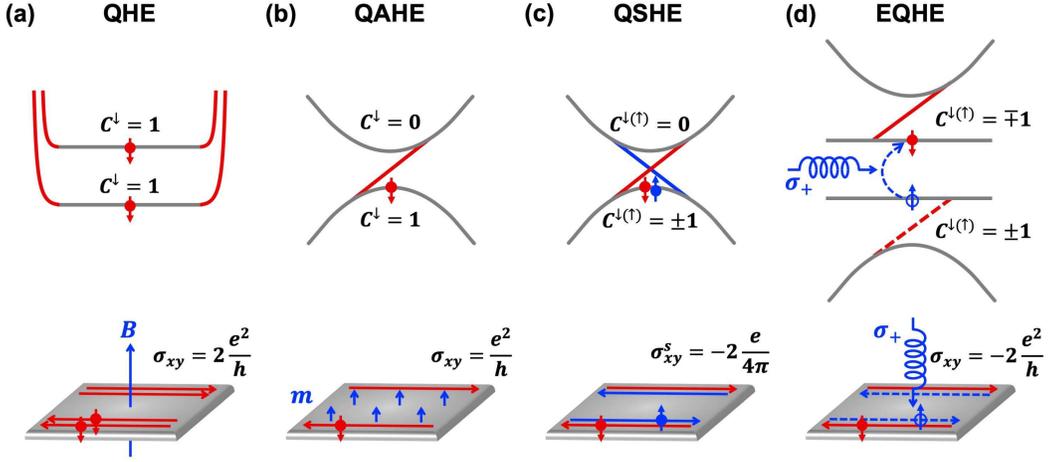}
	\caption{\label{fig:epsart} Schematic illustration comparing four types of quantized charge and spin Hall effects. The upper panel shows the electronic states associated with each effect: (a) QHE featured with LLs and a chiral edge state (red). Each LL carriers a Chern number $C = 1$. (b) QAHE, a Chern insulator state, featured with a topological gap between a dispersive VB of $C = 1$ and a dispersive CB of $C = 0$, and a chiral edge state in the gap (red). (c) QSHE, a topological insulator state, featured with a topological gap between a dispersive VB of spin Chern number $C^s=1/2\,(C^\uparrow-C^\downarrow ) = -1$ and a dispersive CB of $C^s = 0$, a pair of helical edge states in the gap (red and blue lines).  (d) EQHE featured with two enantiomorphic FBs of opposite spin Chern numbers. A right-handed CLP excites a spin-down electron in the upper FB, separated by a topological gap from another CB above with a chiral edge state in the gap (solid red), and a spin-up hole in the lower FBs, separated by a topological gap from another VB below with a chiral edge state in the gap (dashed red). The reverse is true for the left-handed CPL. The lower panel illustrates the quantized charge or spin Hall conductivity carried by topological edge states: (a) $\sigma_{xy} = 2\cdot{e^2}/{h}$ with two filled LLs (see upper panel). (b) $\sigma_{xy} = {e^2}/{h}$ for a Chern insulator with $C = 1$. (c) $\sigma^s_{xy} = -2\cdot{e}/{4\pi}$ for a topological insulator with $C^s = -1$. (d) $\sigma_{xy} = -2\cdot{e^2}/{h}$ for EQHE with an electron-hole pair created in two enantiomorphic FBs of opposite Chern numbers. }
\end{figure*}

QHE, QAHE, and QSHE are all ground-state properties, characterized by a topological gap that is optically inactive because photo-absorption between VB and CB would violate the conservation of orbital angular momentum. Now, let us imagine a band structure that contains two enantiomorphic topological flat bands (FBs) of opposite Chern numbers gapped between two QSH states, as shown in Fig. 1(d). It will allow for a chirality-selective photoexcitation between the two FBs by a circularly polarized light (CPL), enabling an excited-state quantum Hall effect (EQHE). Its photoactivity is similar to that of the valley Hall effect (VHE) \cite{rycerz2007valley,akhmerov2007detection,xiao2007valley,yao2008valley,xiao2012coupled}, but with an important difference in that the excitation between two FBs results in a quantized flat-band Hall conductivity instead of non-quantized dispersive-band Hall conductivity of VHE. In other words, the EQHE consists of a pair of electron- and hole-conducting QAHE or two copies of half QSHE. Effectively the enantiomorphic FBs behave like a condensed-matter version of two filled LLs, giving rise to a $\sigma_{xy}=-2\cdot{e^2}/{h}$, albeit with an electron and a hole occupying two LLs with opposite Chern numbers respectively [Fig. 1(d)]. A more detailed illustration of quantized Hall conductivity for the EQHE is shown in Supplemental Material, Fig. S1.

\begin{figure}
	\centering
	\includegraphics[width=1\columnwidth]{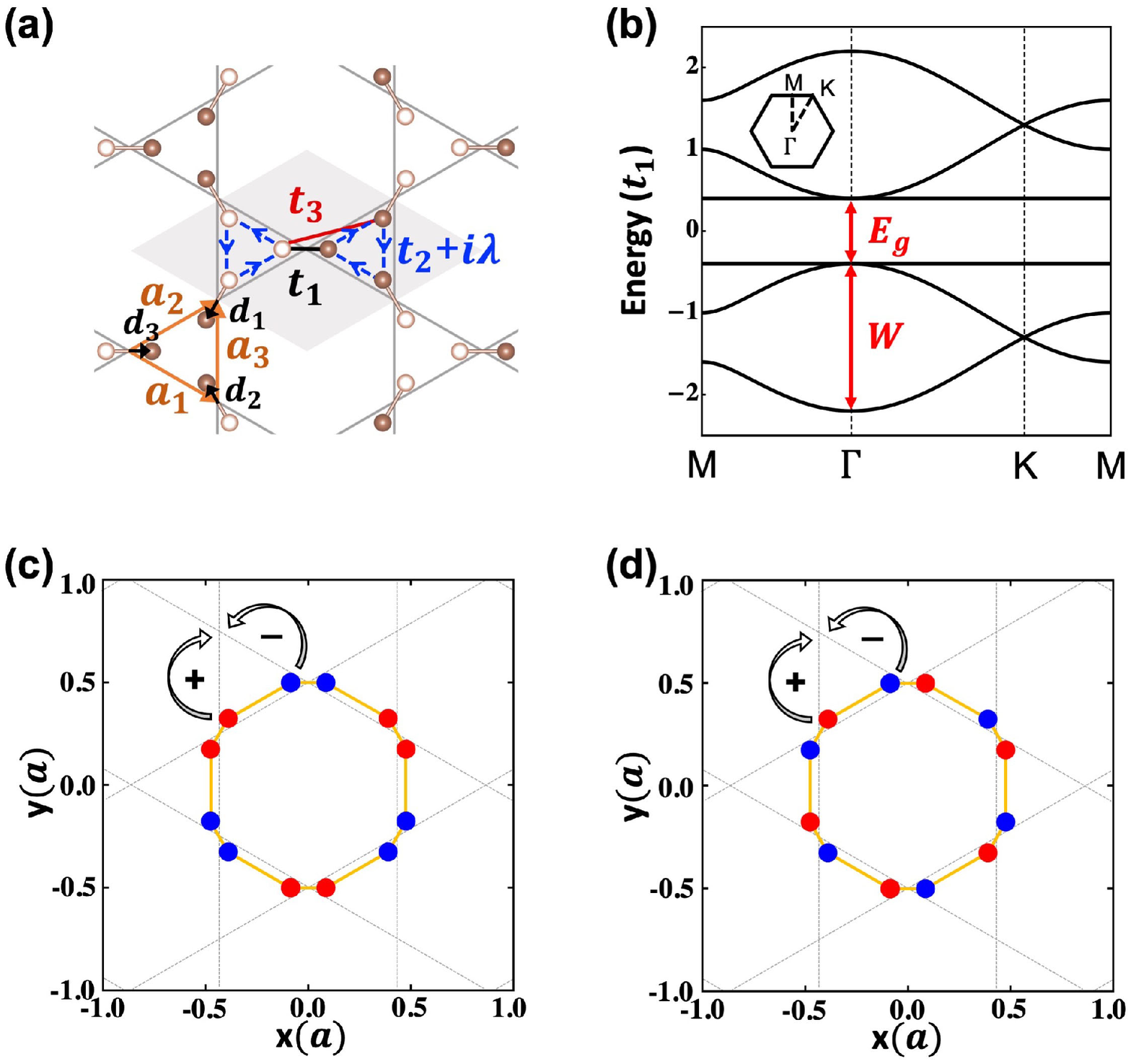}
	\caption{\label{fig:epsart} Creation of the enantiomorphic FBs in a Yin-Yang Kagome lattice. (a) The Yin-Yang Kagome lattice with a dumbbell (two atoms) on every Kagome lattice site. The shaded area indicates the unit cell. $t_1$, $t_2$, and $t_3$ represent the NN intra-dumbbell, 2NN inter-dumbbell and 3NN cross-dumbbell hopping, respectively. $\lambda$ is the SOC strength. (b) The band structure of the Yin-Yang Kagome lattice characterized with two enantiomorphic FBs, obtained with $t_1$, $t_2 = 0$, $t_3 = 0.3t_1$ and $\lambda = 0$. The gap between the two FBs $E_g = |2t_1 - 4t_3 |$; the bandwidth of two sub-Kagome bands $W = |6t_3|$. The inset shows the first Brillouin zone. (c) Distribution of real-space wavefunction of the Yin FB (VB). Red and blue dots represent positive and negative phases of the wavefunction at the position, respectively. (d) Same as (c) for the Yang FB (CB).}
\end{figure}

Next, we discuss the creation of enantiomorphic FBs in a lattice model. It is well-known that a Kagome lattice hosts a FB \cite{mielke1992exact,tang2011high,parameswaran2013wannier,wang2013prediction,zong2016observation}, which lies either above or below two dispersive bands depending on the sign of lattice hopping \cite{ohgushi2000spin,guo2009topological}. Thus, to have two FBs, a natural place to start with is to have two atoms on every Kagome lattice site, creating two sub-Kagome lattices, as shown in Fig. 2(a). A tight-binding Hamiltonian can be written as
\begin{equation}
	\begin{split}
	H &= t_1\sum_{\langle{ij}\rangle\alpha}{c_{i\alpha}^\dagger}c_{j\alpha} + t_2\sum_{\langle\langle{ij}\rangle\rangle\alpha}{c_{i\alpha}^\dagger}c_{j\alpha} + t_3\sum_{\langle\langle\langle{ij}\rangle\rangle\rangle\alpha}{c_{i\alpha}^\dagger}c_{j\alpha}\\
	&+ i\lambda\sum_{\langle\langle{ij}\rangle\rangle\alpha\beta}\frac{2}{\sqrt{3}}(\hat{\textbf{\textit{r}}}_{ij}^1\times\hat{\textbf{\textit{r}}}_{ij}^2)\cdot{\sigma_{\alpha\beta}^z}{c_{i\alpha}^\dagger}c_{j\beta}.
	\end{split}
\end{equation}
The first term represents the nearest-neighbor (NN) intra-dumbbell hopping, $c_{i\alpha}^\dagger$  ($c_{i\alpha}$) is the electron creation (annihilation) operator on site $\textit{i}$ of spin $\alpha$. The second (third) term represents 2NN inter-dumbbell (3NN cross-dumbbell) hopping. The forth term represents spin-orbit coupling (SOC) with a coupling strength $\lambda$, having the cross between two 2NN unit vectors $\hat{\textbf{\textit{r}}}_{ij}^{1,2}$ pointing out from the center of the triangle; ${2}/{\sqrt{3}}\,\vert\hat{\textbf{\textit{r}}}_{ij}^1\times\hat{\textbf{\textit{r}}}_{ij}^2\vert = \pm1$ \cite{kane2005z}. $\sigma_{\alpha\beta}^z$ is the Pauli matrix; $\alpha$ and $\beta$ are spin indices.

However, in its simplest construction considering only the NN ($t_1$) and 2NN hopping ($t_2$), one would obtain a band structure consisting of either one set of Dirac bands plus another set of four bands, or two sets of Kagome bands but both having FB above (or below) the dispersive bands (i.e., the two having the same sign of lattice hopping) for $t_2 < t_1$.

Interestingly, it is found that the trick to obtaining two sets of Kagome bands with opposite signs of hopping is to include a 3NN hopping ($t_3$), which adds a “cross-hopping”, in addition to the intra-hopping ($t_1$), between two sub-Kagome lattices. The condition for obtaining Yin-Yang FBs in terms of these hopping parameters will be discussed later. Here, for simplicity, one can neglect $t_2$, which will not affect the band structure qualitatively in a wide range of parameter space, and use ($t_1$, $t_3$) to construct a beautiful band structure containing a pair of enantiomorphic FBs [Fig. 2(b)]. Specifically, the momentum-space Hamiltonian is expressed as
\begin{subequations}
	\begin{gather}
		H =
		\begin{pmatrix}
			0 & H_0 \\
			H^*_0 & 0
		\end{pmatrix},\\
		H_0 =
		\begin{pmatrix}
			h_0^1 & h_1 & h_2 \\
			h_1 & h_0^2 & h_3\\
			h_2 & h_3 & h_0^3
		\end{pmatrix},
	\end{gather}
\end{subequations}
where $h_0^{1,2,3} = t_1\,e^{-i\boldsymbol{k}\cdot2\boldsymbol{d}_{1,2,3}}$, $h_{1,2,3} = t_3\,cosk_{1,2,3}\,e^{i\boldsymbol{k}\cdot\boldsymbol{d}_{1,2,3}}$, and $k_n = \textbf{\textit{k}}\cdot \textbf{\textit{a}}_n$, $\textbf{\textit{a}}_n$ is the Kagome lattice vector; $d_n$ represents the vector pointing from the Kagome site to the dumbbell atom, as shown in Fig. 2(a). 

Solving Eq. 2, one obtains two sets of distinctive Kagome bands, as shown in Fig. 2(b): the VB below (CB above) the Fermi level (considering a system of half-filling) has the FB above (below) the dispersive bands, resulting from a negative (positive) lattice hopping. Herein we will call this unique lattice model (band structure) the Yin-Yang Kagome lattice (bands), a terminology borrowed from the ancient Chinese Taoism, with Yin and Yang meaning negative and positive respectively, from which a world of intriguing physics will be derived. We note that actually the Yin-Yang Kagome lattice model has been implicitly contained in the so-called hexagonal star lattice studied before \cite{yao2007exact,ruegg2010topological,wen2010interaction,chen2012quantum,chen2012topological}, except that people have viewed the same lattice as having a triangle atomic basis in a hexagonal lattice instead of a diatomic basis in a Kagome lattice and focused on the band structures and physics different from our focus here.

The eigenvalues of the two Yin and Yang FBs are
\begin{equation}
	E_{Y-,Y+} = \pm(t_1-2t_3).
\end{equation}
As shown in Fig. 2(b), both the Yin and Yang Kagome bands have a band width of $|6t_3|$, and the gap between the FBs equals to $|2t_1-4t_3|$. To better understand the physical origin of FBs, their Bloch states are derived as
\begin{equation}
	\begin{split}
		\Psi_{Y-,Y+}^\dagger(\boldsymbol{k})
		=\sum_{i=1}^3(-1)^{i}\,sink_i\,[&\mp\,e^{-i\boldsymbol{k}\cdot\boldsymbol{d}_i}\,c_{\boldsymbol{k},i}^\dagger \\
		&-\,e^{i\boldsymbol{k}\cdot\boldsymbol{d}_i}\,c_{\boldsymbol{k},{i+3}}^\dagger].
	\end{split}
\end{equation}
And Fourier transforming Eq. 4 into real space, one has
\begin{equation}
	\begin{split}
	\Psi_{Y-,Y+}^\dagger(\boldsymbol{R}) & = N\int_{BZ}e^{-i\boldsymbol{k}\cdot\boldsymbol{R}}\Psi_{Y-,Y+}^\dagger(\boldsymbol{k}) \\
	& = \frac{1}{\sqrt{12}}\big\{\sum_{i=1}^6\pm(\mp1)^i c_i^\dagger+\sum_{i=7}^{12}(\mp1)^i c_i^\dagger\big\},
	\end{split}
\end{equation}
where $N$ is a normalization constant and $i$ runs over the twelve vertices of a dodecahedron centered at a chosen position $\textbf{\textit{R}}$, as shown in Figs. 2(c) and 2(d).

Fig. 2(c) and 2(d) show the phase distribution of the real-space wave functions of the Yin and Yang FBs, respectively. They both display an almost perfect pattern of destructive interference (or phase cancellation) expected for a topological FB \cite{zheng2014exotic}. The wave functions are completely localized around a hexagonal plaquette with alternating phases of $(+,-)$, so that lattice hopping out of the plaquette is prohibited [see arrows in Fig. 2(c) and 2(d)]. Moreover, looking at the diatomic basis (dumbbell) on each Kagome lattice site, one sees that for the Yin (Yang) FB the wave functions on the dumbbell are of the same (opposite) phase, indicating a bonding (anti-bonding) state between the two sub-Kagome lattices. The different bonding nature of Yin vs. Yang FB is consistent with their respective negative and positive lattice hopping.

Next, we consider the effect of SOC (the fourth term in Eq. 1). Fig. 3(a) shows the resulting band structure. The SOC opens four topological gaps, and the system is basically a QSH insulator if the Fermi energy lies in one of them. This is further illustrated by the calculated spin Hall conductance \cite{wang2013prediction} showing four quantized plateaus of $\sigma^s_{xy} = -2\cdot{e}/{4\pi}$, in the energy windows of SOC gaps in Fig. 3(b). However, of our interest here is the two isolated FBs that carry opposite Chern numbers in each spin channel, forming an enantiomorphic pair. In Fig. 3(c) we plot Berry curvatures for the two spin-down FBs in the first Brillouin zone (BZ), which look like a mirror image of each other having negative and positive values in the upper and lower FBs respectively. The spin-up FB Berry curvatures are just the opposite of Fig. 3(c). Integration of Berry curvatures over the BZ gives the opposite Chern numbers for the two FBs.

The two enantiomorphic topological FBs of opposite Chern numbers are optically active. Most interestingly, photoexcitation occurs over the whole band, giving rise to a quantized Hall conductivity that equals to Chern number. Also, photoexcitation is spin selective with the left- ($\sigma^-$) and right-handed ($\sigma^+$) CPL to excite the spin-up and -down channel, creating a spin-up and -down electrons in the CB and spin-down and -up holes in the VB, respectively. In Fig. 3(d), we show the calculated inter-FB photo-absorption spectrum of the $\sigma^+$ CPL for the spin-down channel, which shows a strong delta-function-like peak at the gap energy. The same result is obtained for the $\sigma^-$ CPL for the spin-up channel. Consequently, the CPL photoexcitation breaks instantaneously the TRS of a special QSH ground state featured with enantiomorphic FBs, leading to a non-equilibrium EQHE.

\begin{figure}
	\centering
	\includegraphics[width=1\columnwidth]{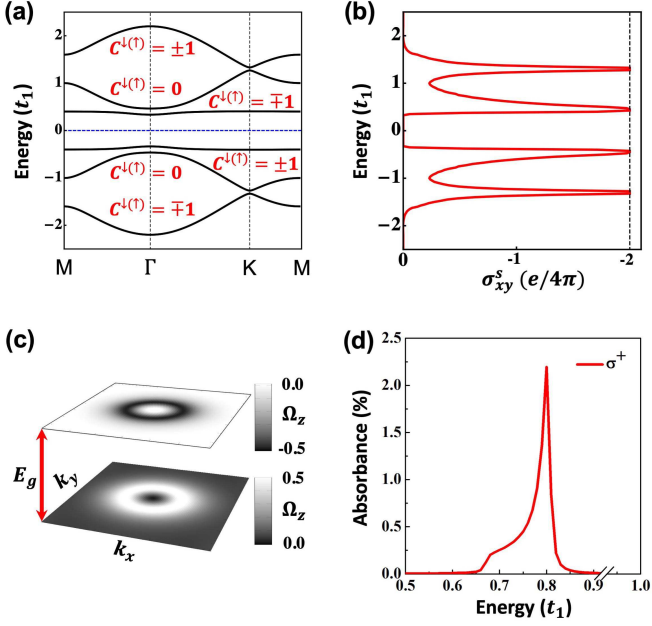}
	\caption{\label{fig:epsart} Topological properties of the Yin-Yang Kagome band. (a) The band structures and Chern numbers with SOC ($\lambda = 0.0375 t_1$). (b) The quantized spin Hall conductivity ($\sigma^s_{xy}$) within the energy window of the SOC gaps. (c) Plots of Berry curvatures ($\Omega_z$) in the first BZ for the two FBs in (a), which are the mirror image of each other due to the opposite Chern number. (d) The inter-FB photo-absorption spectrum of the right-handed CPL ($\sigma^+$) for the spin-down channel. The same is true for the left-handed CPL ($\sigma^-$) for the spin-up channel.}
\end{figure}

It is interesting to compare the EQHE with the VHE \cite{rycerz2007valley,akhmerov2007detection,xiao2007valley,yao2008valley,xiao2012coupled}, as both are photoexcited non-equilibrium Hall effects based on the similar selection rule, to be observed under illumination \cite{mak2014valley}. The key difference is that the former arises from enantiomorphic FBs over the whole BZ leading to a quantized Hall conductivity (see Eq. S1), while the latter from “enantiomorphic valleys” at specific $k$-points leading to a single valley non-quantized Hall conductivity (see Eq. S3). Note that the Berry curvature of FBs shown in Fig. 3(c) has a rather uniform ring-type distribution, which corresponds to the localized real-space distribution of wave function shown in Fig. 2(c) and 2(d). In contrast, the Berry curvature is localized at special $k$-point of valleys in the VHE, corresponding to extended Bloch wave functions in real-space. Consequently, the EQHE has a much stronger intensity, because photo-absorption occurs over the whole band (whole BZ), without the need to break inversion symmetry to create a valley imbalance as for VHE.

The EQHE not only extends Hall physics to excited state with quantized photo-absorption but also offers new applications, such as for making high fidelity topological photodetector and sensor via robust edge photocurrent. Also, the direction of photocurrent can be easily reversed by the handedness of CPL, instead of changing magnetization direction for QAHE, offering a simple switching mechanism for quantum devices. Beyond the EQHE, the enantiomorphic FBs have far-reaching implications from both the fundamental and practical points of view. All the well-known physical effects associated with a single FB, such as ferromagnetism \cite{mielke1992exact,mielke1991ferromagnetism,mielke1991ferromagnetic,tasaki1992ferromagnetism,zhang2010proposed}, Wigner crystallization \cite{wu2007flat,wu2008p}, superconductivity \cite{miyahara2007bcs,kobayashi2016superconductivity}, and fractional QHE \cite{tang2011high,neupert2011fractional}, will also manifest in the two enantiomorphic FBs but with some new twists. Potential applications arising from topological photoactivity includes circular dichroism, lasing, flat-band excitons, and optoelectronic devices. We believe there are much more to be revealed from the Yin-Yang Kagome lattice. This makes the feasibility of its realization in real materials a high order. At last, we will discuss two candidate materials systems to illustrate this possibility.

\begin{figure}
	\centering
	\includegraphics[width=1\columnwidth]{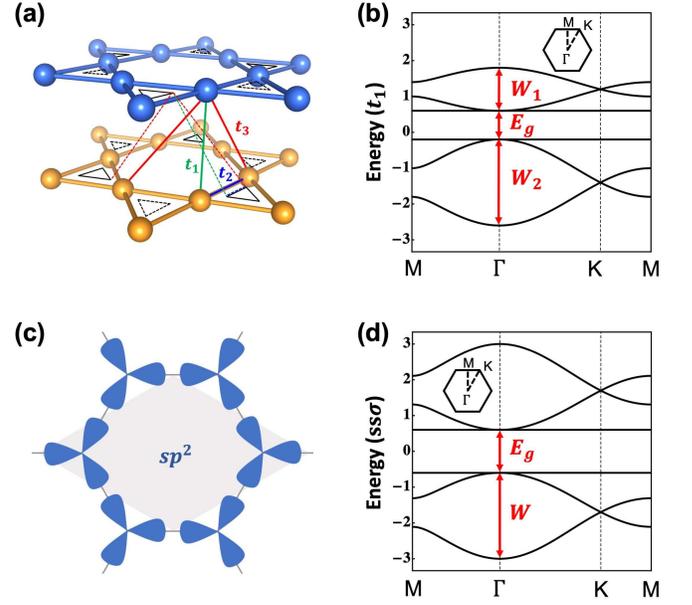}
	\caption{\label{fig:epsart} (a,b) An AA stacking bilayer Kagome lattice (left) exhibiting Ying-Yang Kagome band (right), calculated with $t_2 = 0.1t_1$ and $t_3 = 0.3t_1$. The band gap between the two FBs $E_g = |2t_1 - 4t_3 |$ and the bandwidth $W_1 = 6 |t_2 - t_3 |$ and $W_2 = 6 |t_2 + t_3 |$. (c,d) An $sp^2$ hexagonal lattice (left, the shaded area indicates the unit cell) exhibiting Yin-Yang Kagome bands (right), calculated with  $sp\sigma = -0.8 ss\sigma$, $pp\sigma = -0.4 ss\sigma$, $pp\pi = 0$. The band gap between the two FBs $E_g = 3 pp\sigma$ and the bandwidth $W = 3 |ss\sigma-pp\sigma/2|$. The inset shows the first BZ. }
\end{figure}

The first system is a bilayer Kagome lattice which naturally hosts two sub-Kagome lattices, exhibiting a Yin-Yang Kagome band, as shown in Fig. 4. As illustrated by the indicated hopping patterns, its construction provides an interesting view to understanding the physical origin of Yin-Yang Kagome lattice. By choosing $t_3 > t_2$, one can imagine there are two sub-Kagome lattices formed by cross hopping between the two layers with the top-left (or bottom-left) triangles hopping to the bottom-right (or top-right) triangles, respectively. So, $t_1$ and $t_3$ represent equivalently the intra- and crossing-dumbbell hopping, respectively, shown as the colored dash lines in Fig. 4(a).  It is important to note that bilayer Kagome lattices have been already made experimentally in $\pi$-conjugated nickel-bis(dithiolene) \cite{kambe2013pi}. Density functional theory (DFT) calculations of bilayer nickel-bis(dithiolene) is shown in Fig. S2, whose band structure indeed contains two enantiomorphic FBs. This indicates the hopping condition of $t_3 > t_2$ is satisfied in this lattice (see detailed discussion in Supplemental Material). The second system is the $sp^2$ basis on a hexagonal lattice, which also gives rise to Yin-Yang Kagome bands, as shown in Fig. 4(d). It can be realized in molecular lattices, such as the ones made of triangular graphene flakes. DFT calculations of the patterned graphene-flake superlattice are shown in Fig. S3. With the increasing size of graphene flake (structural motif), the carbon $p_z$ orbitals hybridize into molecular $sp^2$ orbitals on the hexagonal lattice, creating the Yin-Yang FBs [Fig. S3(c,f)]. The details of DFT calculations are presented in the Supplemental Material \footnote{See Supplemental Material at http://link.aps.org/supplemental/xxx, for more details which include Ref.\cite{xiao2007valley,xiao2012coupled,kresse1999ultrasoft,perdew1996generalized,perdew1996generalized,kresse1996efficient,hobbs2000fully}.}. Furthermore, it is worth mentioning that the Yin-Yang Kagome lattice should be experimentally makeable in artificial systems, such as cold-atom lattice \cite{ruostekoski2009optical,jo2012ultracold}, photonic crystal \cite{russell2014hollow}, optical lattice \cite{zhang2019kagome,jiang2019topological}, phononic lattice \cite{xue2019acoustic} and topolectrical circuit \cite{koch2010time}, which opens a number of new topics of study in these fields.

We thank Drs. Zheng Liu, Congjun Wu, Di Xiao, and Chao Zhang for helpful discussions. This work is supported by U.S. DOE-BES (Grant No. DE- FG02-04ER46148).

\nocite{*}

\end{document}